\documentclass[aps,prl,showpacs,notitlepage,nofootinbib,superscriptaddress,floatfix,showkeys,twocolumn,preprintnumbers]{revtex4-1}

\usepackage{blindtext}
\usepackage{hyperref}
\usepackage{amsmath,amssymb}
\usepackage{float}
\usepackage{microtype}
\usepackage{graphicx}
\usepackage{bm}
\usepackage{latexsym}
\usepackage{epsfig}
\usepackage{psfrag}
\usepackage{color}
\usepackage[dvipsnames]{xcolor}
\usepackage{subfigure}
\usepackage[section]{placeins}
\usepackage{comment}

\def\e1i{\epsilon_{1\mathrm{i}}}

\def\dpdm{DPDM\,\,}

\allowdisplaybreaks[1]

\begin{document}

\title{Out of the darkness:\\ probing the inflationary era with dark photon dark matter }

\author{Pierluca Carenza}
\email{pierluca.carenza@fysik.su.se}
\affiliation{Stockholm University and The Oskar Klein Centre for Cosmoparticle Physics, Alba Nova, 10691 Stockholm, Sweden}

\author{Tassia Ferreira}
\email{tassia.ferreira@newcastle.ac.uk}
\affiliation{Department of Physics, University of Oxford, Denys Wilkinson Building, Keble Road, Oxford OX1 3RH, UK}
\affiliation{School of Mathematics, Statistics and Physics, Newcastle University, Herschel Building, NE1 7RU Newcastle-upon-Tyne, UK}

\author{Thong T. Q. Nguyen}
\email{thong.nguyen@fysik.su.se}
\affiliation{Stockholm University and The Oskar Klein Centre for Cosmoparticle Physics, Alba Nova, 10691 Stockholm, Sweden}

\smallskip
\begin{abstract}

A recent hint reported by the TASEH haloscope suggests the possible detection of dark photon dark matter with mass $19.5~{\rm \mu eV}$. Due to their production during inflation, dark photons act as unique messengers from this primordial epoch. We explore the implications that a confirmed detection would have in directly probing the inflationary era for the first time.
To resolve the intrinsic degeneracy between the dark photon mixing parameter  and its fractional relic abundance introduced by haloscope measurements, we motivate a next-generation light-shining-through-a-wall experiment. Combining these two experiences with cosmological data, particularly measurements of the tensor-to-scalar ratio $r$, we propose an interdisciplinary approach to reconstruct dark photon properties.
We delineate a coherent strategy for simultaneously determining the dark photon kinetic coupling, abundance, and properties of the inflationary era.

 \end{abstract}

\maketitle

{\bf \emph{Introduction---}} 
Dark photons (DPs), vector bosons associated with a hidden $U(1)_{X}$ gauge symmetry, have emerged as compelling dark matter (DM) candidates due to their strong theoretical motivation and experimental accessibility~\cite{Arias:2012az, Bertone:2018krk, Cirelli:2024ssz, Bertone:2004pz, Antel:2023hkf, Caputo:2021eaa}. In minimal models, DPs interact with the Standard Model through kinetic mixing with ordinary photons~\cite{Fabbrichesi:2020wbt, Holdom:1985ag, Linden:2024fby, Okun:1982xi}.
\vspace{-0.1cm}
\begin{equation}
    \mathcal{L}\supset -\frac{\epsilon}{2}A_{\mu\nu}^{\prime}F^{\mu\nu}-\frac{m_{A^{\prime}}^{2}}{2}A_{\mu}^{\prime}A^{\prime\mu}\,,
    \vspace{-0.1cm}
\end{equation}
where $F_{\mu\nu}$ and $A^{\prime}_{\mu\nu}$ represent the field strength tensors of the photon and the DP, respectively; $m_{A^{\prime}}$ is the DP mass and $\epsilon$ is a small mixing parameter. This feeble coupling allows us to probe light \dpdm in haloscopes, drawing a compelling parallel with axion searches~\cite{Sikivie:1983ip, Brubaker:2016ktl, ADMX:2018gho, Nguyen:2019xuh, HAYSTAC:2020kwv, CAPP:2020utb, Cervantes:2022epl, Cervantes:2022gtv, McAllister:2022ibe, Cervantes:2022yzp, Schneemann:2023bqc, Nguyen:2024kwy, Linden:2024uph, Nguyen:2025eva, DelaTorreLuque:2025zjt, SHANHE:2023kxz, APEX:2024jxw, He:2024fzj}. In its simplest form, a haloscope comprises a resonant cavity enhancing the DP conversion into photons within a narrow frequency range. A sensitive detector then measures the power output generated during the conversion process.

Haloscopes have the capability of testing the existence of DPs and determining their mass, unveiling the nature of DM. A conclusive signal from a haloscope would revolutionize our understanding of fundamental physics. Recently, in a reanalysis of the data taken by the Taiwan Axion Search Experiment with Haloscope (TASEH) experiment~\cite{TASEH:2022hfm}, a tentative signal, that can be associated with a DP, was found with a local significance of $4.7\sigma$~\cite{Chang:2025ahb}. This signal can be associated with a DP, with mass $m_{A^{\prime}}=19.5~{\rm \mu eV}$ and mixing $\epsilon=2.2\times10^{-15}$, composing the totality of DM. Motivated by this finding, we discuss the implications of a newly discovered DP on early Universe cosmology. In line with the argument of Ref.~\cite{Kitajima:2024jfl}, we disfavour a Higgs origin for such a small DP mass. Thus, the discovery of light \dpdm is a strong indication of the first known particle acquiring mass through the Stueckelberg mechanism~\cite{Stueckelberg:1938hvi}. Under this assumption, the simplest production mechanism for DPs is via inflationary fluctuations~\cite{Graham:2015rva, Kolb:2020fwh}, making it possible to use them as messengers from the inflationary era. We explore the potential of indirectly probing inflation with unprecedented precision through the measurable properties of \dpdm. In particular, we discuss how to determine the reheating temperature $T_{\rm RH}$, at which radiation domination begins, and the Hubble scale at inflation $H_{I}$, thereby locating in time the inflationary era.

Currently, the reheating temperature is constrained to be larger than $\sim 1$~MeV, in order not to spoil Big Bang nucleosynthesis (BBN)~\cite{Giudice:2000ex}, and it is thought to be smaller than the Grand Unification scale $\sim10^{16}$~GeV~\cite{Ross:1981bu}. The inflation scale is currently constrained at $H_{I}<4.8\times10^{13}~{\rm GeV}$~\cite{Paoletti:2022anb} by the non-detection of tensor modes in the Cosmic Microwave Background (CMB)~\cite{Hertzberg:2014sza}. A theoretical lower limit on $H_{I}$ comes from the requirement of inflation to take place before reheating as
\vspace{-0.1cm}
\begin{equation}
    H_{I}\gtrsim 2.8\times10^{9}~{\rm GeV}\left(\frac{T_{\rm RH}}{10^{14}
~{\rm GeV}}\right)^{2}\,.
\label{eq:vinc}
\vspace{-0.1cm}
\end{equation}
This constraint is particularly stringent if $T_{\rm RH}$ is large, and it excludes $T_{\rm RH}\gtrsim 10^{16}$~GeV because of the upper limit on tensor modes. Thus, at the moment, it is safe to say that there is poor understanding of these key parameters for early Universe cosmology. The possibility of probing inflation through DPs is a promising avenue to shed light on the earliest cosmological era.
The goal of this work is to use haloscopes to determine $m_{A^{\prime}}$, and next-generation light-shining-through-a-wall (LSW) experiments to obtain $f_{\rm DM}$ and $\epsilon$, combined with measurements of tensor modes in cosmological analyses of the CMB to constrain with high precision both $H_{I}$ and $T_{\rm RH}$.

\textbf{\emph{Interpretation of the tentative TASEH signal---}} 
Exploring the implications of the discovery of \dpdm is extremely important, particularly for establishing a framework that can be applied once a detection is confirmed. This work is motivated by a signal reported in a reanalysis of TASEH data, which adds urgency to performing such an investigation. The TASEH experiment currently provides the most stringent constraints on axion DM in the $[19.46,\ 19.84]$~$\mu$eV mass range~\cite{TASEH:2022vvu, TASEH:2022noe}. Unlike the axion case, in which haloscopes rely on magnetic fields to facilitate axion-photon conversion, DPs can be detected without the need for such a background field. This allows for a reanalysis of TASEH data to search for DPs~\cite{Chang:2025ahb}.

Haloscopes searching for axions typically use a magnetic field veto to eliminate spurious signals: if a signal remains when the magnetic field is turned off, it is presumed not to originate from axion-photon conversion. However, this criterion does not apply to DPs, which do not require a magnetic field to convert into photons. As a result, a legitimate DP signal might be mistakenly excluded by this veto. This appears to be the case for a signal observed in the $[4.71017,\ 4.71019]$~GHz range, corresponding to $\sim19.5~{\rm \mu eV}$~\cite{Chang:2025ahb}. The signal can be interpreted as a DP with mixing parameter $\epsilon=2.2\times10^{-15}$ with a local significance of $4.7\sigma$. This interpretation assumes an unpolarized DP background, meaning that the DP-induced electric field is randomly oriented, and that DM is completely made up of DPs~\cite{Chang:2025ahb} (see~\cite{Gorghetto:2022sue, Chen:2024vgh} for recent studies on \dpdm polarization). 
An open question is whether the haloscope might have traversed a DM overdensity, regions with densities up to $10^6$ times the average DM density~\cite{Gorghetto:2022sue}. Such an encounter could significantly boost the signal, facilitating the detection of a DP with a much smaller kinetic mixing parameter. In this scenario, a substantial fraction of \dpdm could reside in compact overdense structures, leading to signal variations on timescales of years. These variations would complicate the confirmation of any eventual hint. For the remainder of this work, we proceed under the assumption that the TASEH signal originates from \dpdm distributed with a local DM density of $0.45~{\rm GeV}~{\rm cm}^{-3}$.

\textbf{\emph{A lower bound on $f_{\rm DM}$---}} 
Haloscopes searching for DPs are sensitive to a combination of the mixing parameter $\epsilon$ and the local fraction of DM composed by DPs, $f_{\rm DP}$. Consequently, a null result could either be due to a small mixing parameter $\epsilon$ or to a suppressed \dpdm fraction, $f_{\rm DP} \ll 1$. Conversely, a positive detection constrains only the combination $\epsilon \sqrt{f_{\rm DP}}$, preventing an independent determination of the mixing parameter without additional observational input. To resolve this degeneracy, it is crucial to incorporate complementary constraints from other experimental and observational probes that are sensitive to $\epsilon$ but do not depend on DPs constituting DM today. These include bounds from stellar cooling~\cite{An:2013yua, Redondo:2013lna}, cosmological probes of energy injection or thermalization~\cite{Fradette:2014sza,Vogel:2013raa} and direct detection experiments~\cite{Essig:2013lka,Antel:2023hkf}. These searches provide model-independent limits on $\epsilon$ across a broad mass range, independent of $f_{\rm DP}$, thereby enabling a more precise interpretation of haloscope results.

\begin{figure}[t!]
\includegraphics[width=\linewidth]{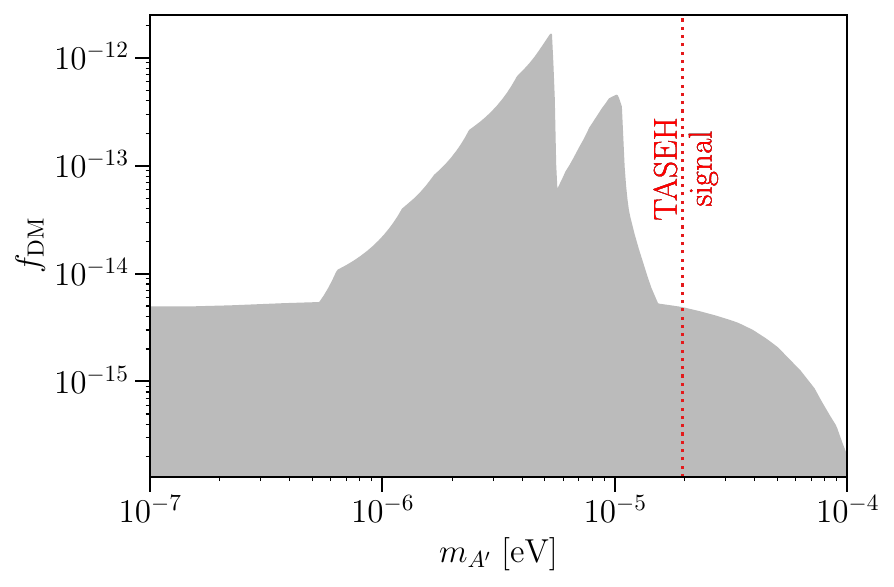}
\vspace{-0.8cm}
\caption{Bound on the \dpdm fraction as a function of mass, constrained by astrophysical~\cite{Caputo:2020bdy, McDermott:2019lch, Chluba:2024wui, Arsenadze:2024ywr} and laboratory data~\cite{Betz:2013dza, Romanenko:2023irv}. The gray region is excluded. Bumps arise from the strong CROWS and DarkSRF limits. The dashed red line marks the TASEH signal mass.}
\vspace{-0.2cm}
\label{fig:fraction}
\end{figure}

For the mass of the TASEH excess, current bounds constrain $\epsilon \lesssim 3.2 \times 10^{-8}$ based on CMB spectral distortions~\cite{Caputo:2020bdy, McDermott:2019lch, Chluba:2024wui, Arsenadze:2024ywr}, and $\epsilon \lesssim 6.4 \times 10^{-8}$ from LSW experiments, such as the CERN Resonant WISP Search (CROWS)~\cite{Betz:2013dza} and the Dark Superconducting Radio Frequency (DarkSRF) cavities search~\cite{Romanenko:2023irv}. These limits allow us to derive a lower bound on the \dpdm fraction required to explain the TASEH signal: $f_{\rm DM} \gtrsim (1.2$–$4.7) \times 10^{-15}$. The range reflects the sensitivity of astrophysical and laboratory constraints. Figure~\ref{fig:fraction} illustrates the constraint on the allowed fraction of DM composed of DPs, based on the most stringent available bounds, as a function of the DP mass. Here, we will show that this lower limit on $f_{\rm DM}$ yields meaningful constraints on key parameters governing the early Universe.

In the event of a confirmed \dpdm detection, significant investments are expected to be directed toward the development of a highly sensitive LSW experiment, aimed at measuring the DP mixing parameter independently of any DM assumptions. A proposal for a next-generation LSW experiment, HyperLSW, is outlined in~\cite{Hoof:2024gfk}, envisioned in the context of an axion discovery scenario. 

\begin{figure}[t!]
\includegraphics[width=\linewidth]{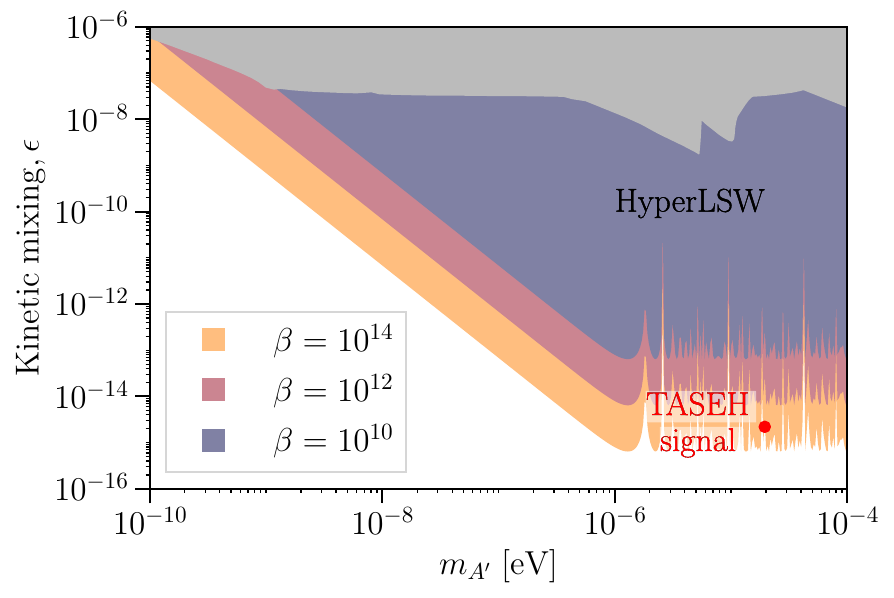}
\vspace{-0.8cm}
\caption{Reach of the HyperLSW experiment for DPs, shown as the colored regions (purple for $\beta=10^{10}$, pink for $\beta=10^{12}$, and orange for $\beta=10^{14}$). The red dot indicates the best fit of the TASEH signal for $f_{\rm DM }=1$ and unpolarized \dpdm. The DP bounds not relying on DM assumptions are shown in grey~\cite{Vinyoles:2015aba, Li:2023vpv, Caputo:2020bdy, McDermott:2019lch, Chluba:2024wui, Arsenadze:2024ywr, Betz:2013dza, Romanenko:2023irv}.}
\vspace{-0.2cm}
\label{fig:HyperLSW}
\end{figure}

The signal generated in an LSW experiment by photon conversion into DP and reconversion is~\cite{Ehret:2010mh, Hoof:2024gfk}
\begin{equation}
    \mathcal{S}=\varepsilon_{\rm eff} \frac{P_{\omega}\Delta t}{\omega}\beta_{g}\beta_{r}P_{A^{\prime}\to\gamma}^{2}\,,
\end{equation}
where $\varepsilon_{\rm eff}$ is the efficiency, assumed to be unity, $P_{\omega}$ is the emitted photon power, $\Delta t$ is the measurement time and $\beta_{g}=\beta_{r}=\beta$ are the boost factors of the generation and regeneration cavities, assumed to be equal, for simplicity. The DP-photon conversion probability, for a DP/photon with energy $\omega$, is~\cite{Ahlers:2007qf}
\begin{equation}
    P_{A^{\prime}\to\gamma}=4\epsilon^{2}\frac{m_{A^{\prime}}^{4}}{M^{4}}\sin^{2}\left(\frac{M^{2}L}{4\omega}\right)\,,
\end{equation}
where $L$ is the path length and $M^{2}=m_{A^{\prime}}^{2}+2\omega^{2}(n-1)$ with $n$ the refraction index, which will be considered equal to unity. 
Compared to an axion LSW experiment, a key advantage is that DPs do not need a magnet for detection, significantly reducing the complexity and cost of the experiment. In Fig.~\ref{fig:HyperLSW}, we present an estimate of the sensitivity of a future DP-specific LSW experiment, based on the following parameters: $\omega=1.3$~GHz analogous to DarkSRF, $L=4$~m, $P_{\omega}=300$~W an order of magnitude larger than CROWS, $\Delta t=5000$~h and $\mathcal{S}\simeq 170$~\cite{Hoof:2024gfk}. Three values of the boost factors are chosen: $\beta=10^{10}$ for the purple region, $\beta=10^{12}$ for the pink region and $\beta=10^{14}$ for the orange region. This experimental setup is compatible with the reach of future LSW experiments~\cite{Ortiz:2020tgs, Miyazaki:2022kxl, Berlin:2022hfx, Antel:2023hkf}. For higher masses, when the mismatch between the wavenumbers of the cavity mode relevant for detection, $k$, and the DP one, $k'$, increases, efficiency decreases~\cite{Betz:2014wie}. This leads to an upper limit on the DP mass that can be probed, above which the detection efficiency drastically drops~\cite{Hoogeveen:1990vq}. This effect is not considered in our simple estimate, and it is not shown in Fig.~\ref{fig:HyperLSW}. We expect that a dedicated design of the detector will allow for probing the mass of the TASEH DP.
The impressive estimated reach will enable verification of the hypothesis that DM is entirely composed of DPs. We are confident that, in a few years, new experimental concepts will make LSW experiments significantly more sensitive~\cite{Burton:2017bxi, An:2025gax}.
Thus, it is reasonable to expect that a confirmation of the TASEH signal leads to the creation of a DP-specific HyperLSW experiment, able to disentangle $f_{\rm DM}$ and $\epsilon$.

{\bf \emph{Inflationary dark photon production---}} 
Various mechanisms for producing \dpdm have been proposed in the literature, broadly classified into thermal~\cite{Redondo:2008ec} and non-thermal scenarios. Non-thermal mechanisms have attracted significant attention due to their ability to naturally accommodate light DP masses. Among these, the misalignment mechanism is particularly notable. In this scenario, the DP field begins coherent oscillations in the early Universe when the Hubble parameter drops below the DP mass scale~\cite{Nelson:2011sf, Arias:2012az}. However, unlike scalar fields, this mechanism is generally inefficient at generating the correct \dpdm abundance unless enhanced by a significant non-minimal coupling to gravity~\cite{Arias:2012az}.
Additional non-thermal DP production channels include parametric resonance of dark Higgs~\cite{Dror:2018pdh} or axions coupled to the DP~\cite{Co:2018lka}, and decay of topological strings, which also assume DP mass generation via a Higgs mechanism~\cite{Long:2019lwl}.

A straightforward and compelling non-thermal production mechanism was proposed in~\cite{Graham:2015rva}, where light \dpdm is generated during inflation via quantum fluctuations. If the DP carries a small but non-zero Stueckelberg mass, its transverse modes undergo quantum fluctuations that become classical and freeze on superhorizon scales. Following inflation, as the Universe becomes radiation dominated, these modes re-enter the horizon and begin to oscillate coherently when $H \sim m_{A^{\prime}}$, behaving as cold DM. This production mechanism exhibits sensitivity to both the inflationary scale and the reheating temperature~\cite{Kolb:2020fwh}. In the event of light \dpdm discovery, this feature becomes a powerful probe of early-Universe cosmology. Motivated by the TASEH tentative signal, this paper adopts precisely this perspective.

\begin{figure*}[t!]
\includegraphics[width=\textwidth]{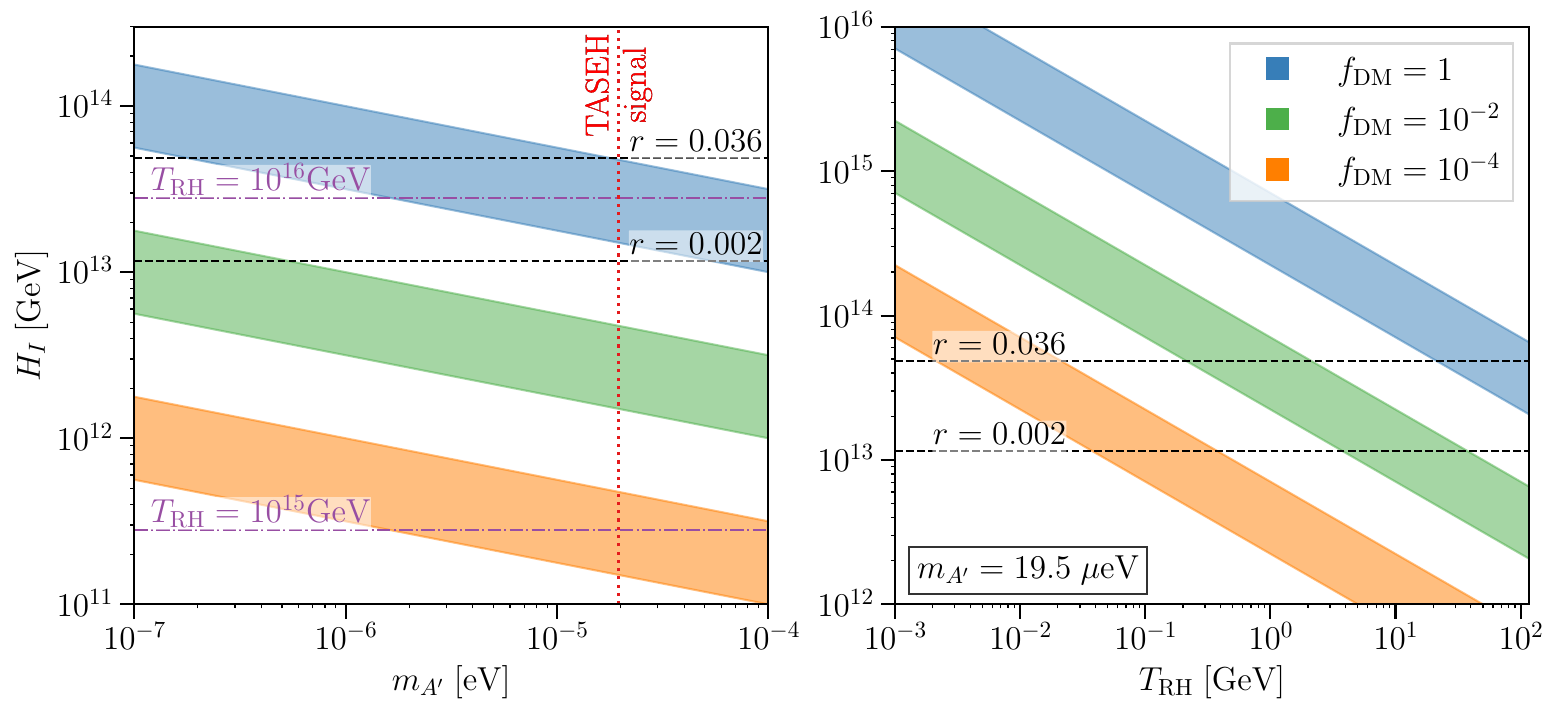}
\vspace{-0.8cm}
\caption{
Hubble scale during inflation $H_I$ as a function of the DP mass $m_{A^{\prime}}$ (left) and reheating temperature $T_{\rm RH}$ (right), yielding \dpdm with fractional abundances $f_{\rm DM}=1$ (blue), $10^{-2}$ (green), and $10^{-4}$ (orange). 
Horizontal black dashed lines show the current upper limit on tensor modes ($r = 0.036$) and projected sensitivity ($r = 0.002$). 
Shaded bands reflect mild dependence on inflationary models. 
\textit{Left:} $H_I$ has a minimum set by $T_{\rm RH}$, shown as horizontal purple dashed lines: $H_I \gtrsim 2.8 \times 10^{13}$ GeV for $T_{\rm RH} = 10^{16}$ GeV and $H_I \gtrsim 2.8 \times 10^{11}$ GeV for $10^{15}$ GeV.
The vertical red line marks the TASEH signal mass. 
\textit{Right:} 
Lower bound on $T_{\rm RH}$ is given by BBN, and its upper bound is calculated for $m_{A^{\prime}} = 19.5~\mu$eV (TASEH). A tensor mode detection in this scenario provides a relation between $T_{\rm RH}$ and $f_{\rm DM}$.}
\vspace{-0.3cm}
\label{fig:combined}
\end{figure*}

{\bf \emph{Constraining early Universe parameters---}} Due to the nature of the DP production mechanism, inflationary imprints are left in the DP relic density today. In order to make this statement quantitative, we recall that the energy density stored in the relic DP field is given by~\cite{Kolb:2020fwh}
\begin{widetext}
\vspace{-0.2cm}
\begin{equation}
\vspace{-0.1cm}
\frac{\Omega \,h^{2}}{0.12}=f_{\rm DM}=\begin{cases}
     (1-10)\times\left(\frac{H_{I}}{10^{11}{\rm GeV}}\right)^{2} \left(\frac{T_{\rm RH}}{5\times10^{7}{\rm GeV}}\right)\left(1-\frac{2}{5}\frac{m_{A^{\prime}}}{H_{I}}\right)&\quad\quad T_{\rm RH}<8.4\times10^{8}{\rm GeV}\left(\frac{m_{A^{\prime}}}{\rm GeV}\right)^{1/2}\,,\\
     (1-10)\times\left(\frac{H_{I}}{10^{14}{\rm GeV}}\right)^{2} \left(\frac{m_{A^{\prime}}}{10^{-6}{\rm eV}}\right)^{1/2}&\quad\quad T_{\rm RH}>8.4\times10^{8}{\rm GeV}\left(\frac{m_{A^{\prime}}}{\rm GeV}\right)^{1/2}\,,\\
\end{cases}
\label{eq:relic} 
\end{equation}
where the weak dependence on the inflationary model is encoded in the numerical prefactor. The dependence of the DP relic density on $H_{I}$ and $T_{\rm RH}$ allows us to probe the cosmology through \dpdm properties. Eq.~\eqref{eq:relic} shows that ${\rm \mu}$eV-scale DPs can constitute the totality of DM provided inflation occurs at a sufficiently high energy scale. Indeed, this relation can be inverted to extract the Hubble scale at inflation for a given DP mass and DM fraction
\vspace{-0.1cm}
    \begin{equation}
    H_{I}=\begin{cases}
        (1.5-4.8)\times10^{13}~{\rm GeV}\sqrt{f_{\rm DM}}\left(\frac{m_{A^{\prime}}}{19.5~{\rm\mu eV}}\right)^{-1/4}\,&\quad\quad T_{\rm RH}>2.6\times10^{4}\left(\frac{m_{A^{\prime}}}{\rm eV}\right)^{1/2}{\rm GeV}\,,\\
        (0.7-2.2)\times 10^{11}~{\rm GeV}\left(\frac{T_{\rm RH}}{10^{7}~{\rm GeV}}\right)^{-1/2}\sqrt{f_{\rm DM}}\,&\quad\quad T_{\rm RH}<2.6\times10^{4}\left(\frac{m_{A^{\prime}}}{\rm eV}\right)^{1/2}{\rm GeV}\,,\\
    \end{cases}
    \vspace{-0.2cm}
    \label{eq:HI}
\end{equation}
\end{widetext}
which should be consistent with Eq.~\eqref{eq:vinc}. Locating inflation in time by determining $H_{I}$ can be achieved by searching for tensor modes in the CMB~\cite{Hertzberg:2014sza, Starobinsky:1979ty, Seljak:1996gy, Kamionkowski:1996zd}: the higher the scale of inflation, the stronger the signal.

An analysis of BICEP2/{\it Keck} Array and {\it Planck} data~\cite{BICEP2:2015nss} sets a constraint on tensor modes, in particular on the tensor-to-scalar ratio defined as~\cite{BICEP2:2015xme, BICEP2:2018kqh, BICEP:2021xfz} 
\begin{equation}
    r(k) = \frac{{\mathcal P}_{\rm t}(k)}{{\mathcal P}_{\mathcal R}(k)}\,,
\label{ratio}
\end{equation}
where ${\mathcal P}_{\rm t}(k)$ and ${\mathcal P}_{\mathcal R}(k)$ are the tensor and scalar power spectra, respectively. Notice that $r$ is scale-dependent and it is measured at some pivot scale, typically chosen to be $k_*=0.05\,$Mpc$^{-1}$. The constraint is $r < 0.035$~\cite{Paoletti:2022anb} at the 95\% C.L. (see also the comparable LIGO/VIRGO bound~\cite{Planck:2018jri, Meerburg:2015zua, LIGOScientific:2016jlg, Cabass:2015jwe}). Combining this upper limit with the measurement of the scalar amplitude $A_{s}$, it is possible to constrain $H_{I}$ since $A_{s}\sim H_{I}^{2}/r$~\cite{Hertzberg:2014sza, Planck:2018jri}. Indeed, the current constraint on $H_{I}$ reads~\cite{Paoletti:2022anb} 
\begin{equation}
H_I < 4.8 \times 10^{13}~{\rm GeV}\,.
\end{equation}
In the next years, there will be several experiments~\cite{SimonsObservatory:2018koc, CMB-S4:2016ple}, such as LiteBIRD~\cite{LiteBIRD:2022cnt}, probing the tensor-to-scalar ratio down to $\sim\mathcal{O}(0.001)$~\cite{LiteBIRD:2022cnt, SimonsObservatory:2018koc, CMB-S4:2020lpa}. 

Currently, the only way of probing the Hubble scale at inflation, $H_{I}$ is by measuring the tensor modes, $r$, with the caveat of providing limited information on the reheating temperature $T_{\rm RH}$. 
On the other hand, the \dpdm detection gives a well-determined mass, which, by assuming a scenario with $T_{\rm RH}\gtrsim200$~GeV, can be used to obtain $H_{I}$ corresponding to a \dpdm fraction, $f_{\rm DM}$. Then, we can predict the corresponding $r$ and compare it with observational constraints and forecasts. This is shown in the left panel of Fig.~\ref{fig:combined}, where the results of Eq.~\eqref{eq:HI} give an inferred value for $H_{I}$ as a function of the \dpdm mass and for various $f_{\rm DM}$. The mass of the TASEH signal is highlighted as a red vertical dotted line. 
By combining both, in a large temperature reheating scenario, it is possible to determine $f_{\rm DM}$ and discriminate among different inflationary models. In the case of low temperature reheating, we can easily obtain the value of $T_{\rm RH}$ for a given $f_{\rm DM}$.

If the totality of DM is composed by DPs with the mass given by the TASEH signal, the tensor modes are indicated by the blue band, whose thickness accounts for a mild model-dependence on the inflationary mechanism. In this case, the tensor modes are just below the sensitivity of current experiments (the $r=0.036$ dashed line) and detectable by next-generation experiments (the $r=0.002$ dashed line). A confirmation of \dpdm detection in TASEH opens the exciting opportunity to independently probe $r$, thus enforcing an upper limit on the reheating temperature $T_{\rm RH}$. As shown in the left panel of Fig.~\ref{fig:combined}, a signal slightly above $r\sim 0.002$, implies that $T_{\rm RH}$ has to be lower than $10^{16}$~GeV, as indicated by the horizontal purple line. The advantage of a joint detection of tensor modes and \dpdm is therefore clear, as it would provide significant constraints on inflationary models. 

From the left panel of Fig.~\ref{fig:combined}, we note that a fraction $f_{\rm DM}\ll1$ implies a low-scale inflation, making the detection of tensor modes difficult.
In this scenario, the existence of \dpdm helps in setting a lower bound on $H_{I}$. In particular, when considering the smallest \dpdm fraction allowed by current bounds $f_{\rm DM}=(1.2-4.7)\times10^{-15}$ (see Fig.~\ref{fig:fraction}), corresponding to the TASEH signal mass, we obtain
\begin{widetext}
\vspace{-0.1cm}
\begin{equation}
\vspace{-0.2cm}
H_{I}\gtrsim\begin{cases}
    (0.5-3.3)\times10^{6}~{\rm GeV}\quad\quad & 117~{\rm GeV}\lesssim T_{\rm RH}\lesssim10^{12}~{\rm GeV}\,,\\
    (0.7-4.8)\times 10^{7}~{\rm GeV}\left(\frac{T_{\rm RH}}{1~{\rm GeV}}\right)^{-1/2}\quad\quad & 1~{\rm MeV}\lesssim T_{\rm RH}\lesssim117~{\rm GeV}\,.\\
\end{cases}
\end{equation}
\end{widetext}
Enhancing the minimum bound on $f_{\rm DM}$ by a factor of $10^{6}$ through future laboratory experiments like HyperLSW, would lead to a constraint on the minimum $H_{I}$ that is stronger by a factor of $10^{3}$. Thus, even in a pessimistic scenario, precious information on inflation is inferred through \dpdm.

In the case that $f_{\rm DM}$ is not much smaller than unity, a tensor mode detection is possible and would point towards a low-temperature reheating, as shown in the right panel of Fig.~\ref{fig:combined}. 
Here, $H_{I}$ needed to produce a given fraction of DM as DPs is shown as a function of $T_{\rm RH}$, for the mass of the TASEH signal.
Even without a determination of $f_{\rm DM}$, measuring $r$ and knowing the nature of DM will impose a relation between $H_{I}$ and $T_{\rm RH}$.

\textbf{\emph{Conclusions---}} 
The discovery of light \dpdm in haloscopes introduces a new and exciting era for particle physics and cosmology, offering a unique window into the early Universe. Haloscope experiments inherently exhibit a degeneracy between the kinetic mixing parameter, $\epsilon$, and the \dpdm fraction, $f_{\rm DM}$. 
Here, we demonstrated how combining haloscopes and LSW experiments breaks this degeneracy and establishes a lower bound on $f_{\rm DM}$. 
Determining the value of $f_{\rm DM}$ helps in obtaining concrete predictions for the inflationary Hubble scale, $H_I$, and thus for the associated primordial tensor modes.

We expect future experiments measuring the tensor-to-scalar ratio $r$ through CMB polarization, such as those by the Simons Observatory~\cite{SimonsObservatory:2018koc}, the Large Scale Polarization Explorer~\cite{LSPE:2020uos}, LiteBIRD~\cite{LiteBIRD:2022cnt}, and CMB-S4~\cite{CMB-S4:2016ple}, to detect a signal if DM consists entirely of DPs with the mass suggested by TASEH. Moreover, detectable tensor modes would also emerge if DPs constitute only a subdominant fraction of DM, provided the Universe underwent low-temperature reheating, with $T_{\rm RH}$ below a few hundred GeV. Within realistic timescales, new measurements of $r$ could appear around the same time as a confirmed \dpdm detection. 
Although determining the value of $f_{\rm DM}$ may not be currently feasible, haloscope observations and cosmological data already provide complementary constraints on the relation between the reheating temperature $T_{\rm RH}$ and the inflationary scale $H_I$. This provides valuable insight beyond our current limits of understanding.

Our discussion applies broadly to any \dpdm detected through haloscope experiments in the mass range $[10^{-7}$, $10^{-4}]$~eV. For even lighter DPs, current bounds on primordial tensor modes already rule out an inflationary production scenario for $f_{\rm DM} = 1$, unless the reheating temperature barely exceeds the MeV scale. On the other hand, for heavier DPs beyond this range, inflationary production would require a very low inflationary Hubble scale $H_I$, resulting in highly suppressed tensor modes.

For a confirmed \dpdm detection, the scenario we explored in this work offers a low-risk high-reward opportunity, as it leverages data from experiments that are either already operational or currently under construction. From this perspective, the proposed HyperLSW experiment, despite its technological challenges, stands out as a highly compelling and low-risk endeavour, given the immense scientific return of creating the brightest DP factory.
Beyond the profound implications of discovering DM and identifying a new particle, the detection of light  \dpdm in haloscopes would mark a pivotal milestone. It would enable us to use DPs as probes of the early Universe, revealing insights into physics at energy scales approaching the Planck scale and constraining cosmological models.

\textbf{\emph{Acknowledgements---}}
We thank Tim Linden for comments on the draft. PC and TTQN are supported by the Swedish Research Council under contract 2022-04283.
TF is supported by a Royal Society Newton International Fellowship and the United Kingdom Science and Technology Facilities Council (STFC) under grant number ST/Y002652/1.
This article/publication is based on the work from COST Action COSMIC WISPers CA21106, supported by COST (European Cooperation in Science and Technology).


\bibliographystyle{apsrev4-1}
\bibliography{references.bib}

\end{document}